\documentclass[sigconf]{acmart}

\usepackage{graphicx} 
\usepackage{booktabs}
\usepackage{multirow}
\usepackage{enumitem}
\AtBeginDocument{%
  }

\setcopyright{none}
\renewcommand\footnotetextcopyrightpermission[1]{}
\acmConference[WSDM '27]{The 20th ACM International Conference on Web Search and Data Mining}{February 15--19, 2027}{Hong Kong}
\acmYear{2027}
\copyrightyear{2027}

\begin{document}

\title{Cluster-Guided Decoding with Trust-Region Constraints for Personalized Recommendation}

\author{Gezheng Wen}
\affiliation{%
 \institution{Amazon}
 \country{USA}
}
\email{wengz@amazon.com}

\author{Chao Shen}
\affiliation{%
 \institution{Amazon}
 \country{USA}
}
\email{shencha@amazon.com}

\author{Raviteja Anantha}
\affiliation{%
 \institution{Amazon}
 \country{USA}
}
\email{ravantha@amazon.com}

\author{Tan Shen Cliff Chung}
\affiliation{%
 \institution{Amazon}
 \country{USA}
}
\email{ctschung@amazon.com}

\author{Layne C. Price}
\affiliation{%
 \institution{Amazon}
 \country{USA}
}
\email{prilayne@amazon.com}

\renewcommand{\shortauthors}{Wen et al.}

\begin{abstract}
Most recommenders score each item in the catalog separately, and do not explicitly model which semantic region of the catalog a recommendation belongs to. Models expressive enough to offer that control are often too slow to run over a full catalog for online inference. In this paper, we present Cluster-Guided Generative Reconstruction (CG2R), which learns a hierarchy of semantic clusters and uses a cluster-level prior to steer item ranking for personalized recommendations. Unlike existing systems that only use clusters to pre-select candidates, CG2R adds the cluster prior directly to the item scoring rule, so that the learned cluster structure shapes the ranking itself. Because the prior is assembled from cluster centroids rather than per-item statistics, it remains defined even for cold-start users when no history is available. Across the three benchmark datasets evaluated, CG2R outperforms classical, neural, and LLM-based baselines in terms of Recall and nDCG, with its performance gains widening as the catalog grows sparser. In addition, CG2R serves recommendations in milliseconds, suggesting potential as low-latency first-stage retrieval.

\end{abstract}



\maketitle

\section{Introduction}

A real-world recommender has to balance three demands at once: it must be efficient enough to run over a large item catalog, personalized to individual users, and controllable in what it recommends. Classical collaborative filtering, such as Matrix Factorization (MF)~\cite{koren2009matrix} and Probabilistic Matrix Factorization (PMF)~\cite{mnih2007probabilistic}, retrieves quickly over large catalogs but relies primarily on user–item interactions. Without side information or specialized extensions, they typically handle cold-start users and items poorly. More expressive neural and sequential models, including NeuMF~\cite{he2017neural}, SASRec~\cite{kang2018self}, BERT4Rec~\cite{sun2019bert4rec}, and LightGCN~\cite{he2020lightgcn}, improve representation learning and temporal modeling, but they still score each item on its own and carry no explicit notion of the semantic region a recommendation belongs to. Large language model (LLM)-based recommenders~\cite{netflix2025foundation,openai2025gpt52} can model user preferences flexibly, but their inference cost and latency block them from serving as first-stage retrieval over a large catalog. Retrieval-augmented generation (RAG)~\cite{lewis2020retrieval} and semantic clustering~\cite{hou2024blair} improve grounding and coherence by treating retrieval as a preprocessing step that narrows the candidate pool, but leaves the ranking function itself blind to semantic structures. Overall, existing recommendation approaches tend to sit at one of two extremes: they scale well but give no control over which semantic region of the catalog recommendations come from, or they reason expressively but too slowly for online inference under a tight latency budget.

We argue that such limitations could be overcome by making semantic cluster structure part of the recommender itself. Specifically, instead of scoring $N$ catalog items independently, we can learn a hierarchy of semantic clusters and a cluster-level prior to steer item ranking, where the clusters specify an explicit semantic region of the catalog to stay within, and the prior is trained with $K$ cluster centroids ($K \ll N$).

We propose \textbf{CG2R} (Cluster-Guided Generative Reconstruction), which consists of a three-phase framework: 1) the cluster retrieval phase learns a hierarchy of semantic item clusters and the user-cluster affinities from users' interaction histories; 2) the contextual expansion phase refines the cluster-level prior using the clusters each user has previously interacted with; 3) the constrained decoding phase applies the learned user-specific prior as a trust region over the catalog, similar to the idea of trust-region policy optimization~\cite{schulman2015trust,schulman2017proximal}. The key difference between CG2R and prior cluster-based systems lies in where the cluster guidance acts: those systems only use clusters to preselect a candidate pool, or to regularize training, while CG2R adds the learned cluster prior directly to each item's semantic-relevance score. The cluster structure therefore influences the final ranking itself.

We evaluate CG2R on three public benchmarks that vary widely in interaction density and catalog size, against classical, neural, and LLM-based baselines. CG2R is the strongest method on all three e.g., improving Recall@10 by 37\%-127\% over the best baseline. In terms of latency, CG2R responds to requests in milliseconds where an LLM-based recommender needs tens of seconds. We make the following contributions. (1) We propose cluster-guided decoding, a scoring rule that adds a cluster prior to semantic relevance, so that learned cluster structure enters the ranking function directly. (2) We construct the prior hierarchically and contextually, combining user-cluster affinity with the clusters in a user's history, so that cold-start users with no interaction history fall back to the population-level user-item affinity. (3) We express the trust region at two points in the pipeline, as an additive bias inside the decoder and as a divergence penalty on the objective function. (4) We make the cluster guidance strength a learned scalar, so that how much the model relies on cluster guidance is determined during training rather than empirically chosen.

\section{Methodology}

CG2R operates in three phases. Phase~1 learns the cluster hierarchy and the item representation that seed the prior $q$ (Section~\ref{subsec:phase1_details}); Phase~2 assembles the user-specific prior $q$ from the user's history $\mathcal{H}_u$ (Section~\ref{subsec:phase2_details}); and Phase~3 applies $q$ as a trust region during scoring (Section~\ref{subsec:phase3_details}).

\subsection{Overview}
\label{subsec:overview}

Let $\mathcal{V}$ be a catalog of $N$ items, and let user $u$ have an interaction history $\mathcal{H}_u = (i_1, \ldots, i_T)$ ordered in time 
where $i\_j$ denotes the item at position $j$ in the user’s chronological interaction history. Given $\mathcal{H}_u$, the task is to rank the catalog by the likelihood of each item $i$ in $\mathcal{V}$ being the user’s next interaction and return the top-$k$ items as recommendations.

Rather than scoring each item $i$ purely on its similarity to the user $u$, CG2R organizes the catalog $\mathcal{V}$ into a hierarchy of semantic clusters and uses a cluster-level prior to steer the scoring. It consists of two trained components: 1) a fused item representation $f_i$ that is both semantic and cluster-aware, and 2) a sequential generator that consumes $\mathcal{H}_u$ and predicts the next-item representation $\hat{x}_u$. The personalized score of item $i$ for user $u$ is
\begin{align}
\label{eq:cg2r_decode}
s(i \mid u) &= \underbrace{\frac{1}{\tau} \langle \hat{x}_u,\, f_i \rangle}_{\text{semantic relevance}} \;+\; \underbrace{\gamma \, \log q(i \mid u, \mathcal{H}_u)}_{\text{cluster guidance}},
\end{align}
where $\langle\cdot,\cdot\rangle$ denotes the dot product, $\tau$ is a temperature constant, $\gamma$ is a learned scalar controlling how much the cluster guidance is applied, and $q$ is the cluster-to-item prior defined in Equation~\eqref{eq:item_prior}. The first term distinguishes items by personalized relevance, while the second favors items belonging to user-relevant semantic clusters. The rest of this section describes how CG2R builds and applies these components.

\subsection{CG2R Framework}
\label{subsec:framework}

\subsubsection{Phase 1: Cluster Learning and Item Representation}
\label{subsec:phase1_details}

This phase builds the item representation and the cluster structure that CG2R relies on.

\textbf{Continuous Item Embedding:} Starting from a semantic representation of each item, we apply the Qwen3-Embedding-4B model ~\cite{zhang2025qwen3embedding} to each item $i$. Using its Matryoshka Representation Learning ~\cite{kusupati2022matryoshka} support, we then truncate the output to its leading $1024$ dimensions to obtain a vector $x_i$.

\textbf{Hierarchical Item Encoding:} To organize items into hierarchical clusters, we use $x_i$ to train a Residual Quantization Variational Autoencoder (RQ-VAE)~\cite{lee2022autoregressive}, where its encoder decomposes $x_i$ into $L$ levels of discrete codes, each level quantizing the residual left by the levels above it, and its decoder reconstructs the item from the hierarchical codes:
\begin{align}
r_\phi(z_{i,1:L}\mid i) &= \prod_{l=1}^L r_\phi(z_{i,l}\mid i, z_{i,1:l-1}),\\
p_\theta(i\mid z_{i,1:L}) &= \text{softmax}\!\left(W_{\text{dec}} \cdot [e_1(z_{i,1}); \ldots; e_L(z_{i,L})]\right).
\end{align}
Here, $z_{i,l}$ is item $i$'s discrete code at level $l$, $e_l(z_{i,l})$ is its RQ-VAE codebook embedding, $W_{\mathrm{dec}}$ is the decoder projection, and $\phi$ and $\theta$ parameterize the RQ-VAE encoder and decoder.
The code prefix of length $l$ names the item's cluster at level $l$, and these codes induce the cluster hierarchy $\{C^{(1)}, C^{(2)}, \ldots, C^{(L)}\}$ used throughout the approach. We treat the codes as the item's semantic ID~\cite{tay2022transformer, rajput2023recommender}.

\textbf{Fused Item Embedding:} We then combine the continuous embedding $x_i$ and the discrete codes into the single representation of $i$ for scoring,
\begin{align}
\label{eq:fused}
f_i &= W_p\, x_i \;+\; \sum_{l=1}^{L} \text{emb}_l\bigl(z_{i,l}\bigr),
\end{align}
where $W_p$ projects $x_i$ into the scoring space and $\text{emb}_l(\cdot)$ is the learned codebook embedding for level $l$.

\textbf{Cluster Statistics:} Let $\mathcal{C}_{l,c}$ denote cluster $c$ at level $l$. For every cluster, we store a centroid $\mu^{(l)}_c$, computed as the mean of its member items' fused embeddings, normalized to the unit sphere:
\begin{align}
\label{eq:centroid}
\mu^{(l)}_c &= \frac{1}{|\mathcal{C}_{l,c}|}\sum_{i \in \mathcal{C}_{l,c}} \tilde{f}_i,
\end{align}
Each centroid summarizes its cluster as a single point, giving Phase~2 a compact target to compare a user against when it retrieves clusters.

\subsubsection{Phase 2: Contextual Expansion Learning}
\label{subsec:phase2_details}

Phase~2 uses the artifacts from Phase~1 and each user's interaction history to build the user-specific prior $q$ in Equation~\eqref{eq:cg2r_decode} in three steps: predict a next-item representation, estimate user-cluster affinities and refine them using the user’s history, and spread the resulting cluster mass over items.

\textbf{Sequential Generator:} We encode the user history $\mathcal{H}_u$ with a causal Transformer over the fused embeddings of its items ($f_{i_1}, \ldots, f_{i_T}$), whose attention layers take an additive cluster bias derived from the cluster centroids of the history items. Reading out the final position gives the predicted next-item representation $\hat{x}_u$. The generator is trained end-to-end with the rest of the model using the objective described later in Phase~3 (Section~\ref{subsec:phase3_details}).

\textbf{Cluster Retrieval:} We score how well each cluster fits the user by comparing $\hat{x}_u$ against the cluster centroids. At each level $l$, a softmax over centroid similarities gives a distribution over that level's clusters,
\begin{align}
\label{eq:cluster_affinity}
p_1(c \mid u) &= \frac{\exp\!\big(\langle \hat{x}_u, \mu^{(l)}_c \rangle / \tau_c\big)}{\sum_{c'} \exp\!\big(\langle \hat{x}_u, \mu^{(l)}_{c'} \rangle / \tau_c\big)},
\end{align}
where $\tau_c$ is a temperature. This gives an initial estimate of the user's cluster preferences from $\hat{x}_u$ alone, which the next step refines using the history.

\textbf{Contextual Expansion:} We then refine the affinity $p_1(c \mid u)$ with the clusters the user has previously interacted with. Letting $\hat{h}_u(c)$ denote the empirical distribution of clusters over the items in $\mathcal{H}_u$, the expanded cluster distribution blends the two,
\begin{align}
\label{eq:phase2_blend}
p(c \mid u, \mathcal{H}_u) &= \bigl(1 - w_2 \mathbf{1}[\mathcal{H}_u \neq \emptyset]\bigr)\, p_1(c \mid u) \notag \\&\quad + w_2 \mathbf{1}[\mathcal{H}_u \neq \emptyset]\, \hat{h}_u(c),
\end{align}
where $w_2 \in [0,1]$ sets how much the history reshapes the affinity and $\mathbf{1}[\cdot]$ is the indicator function. For cold start, if a user has no history the second term drops out and $p(c \mid u, \mathcal{H}_u)$ reduces to the retrieval affinity $p_1(c \mid u)$.

\textbf{Cluster-to-Item Prior:} Finally, we turn this cluster distribution $p(c \mid u, \mathcal{H}_u)$ into a prior $q$ over all items in $\mathcal{V}$ by spreading each cluster's mass uniformly across its members,
\begin{align}
\label{eq:item_prior}
q(i \mid u, \mathcal{H}_u) &\propto \sum_{l} \omega_l \, \frac{p(c_l(i) \mid u, \mathcal{H}_u)}{|\mathcal{C}_{l,c_l(i)}|},
\end{align}

where the level weights $\omega_l$ are learned by a softmax over per-level logits. This $q$ is the cluster guidance term of Equation~\eqref{eq:cg2r_decode}, which Phase~3 applies.

\subsubsection{Phase 3: Constrained Decoding}
\label{subsec:phase3_details}

Phase~3 applies the prior $q$ as a soft trust region. The prior $q$ enters the scoring rule in Equation~\eqref{eq:cg2r_decode} as the additive term $\gamma \log q$, which shifts scores toward the user's likely clusters while keeping every item in $\mathcal{V}$ a candidate. Because the score stays differentiable everywhere, the same scoring rule is used at inference and as the training objective below.

\textbf{Training Objective:} We treat the cluster-guided scores $s(i \mid u)$ as logits, so that training propagates gradient through the generator, the fused embeddings, and the cluster guidance strength $\gamma$ while the trained cluster hierarchy from Phase~1 stays fixed. Normalizing the scores $s(i \mid u)$ over the catalog $\mathcal{V}$ gives the next-item distribution for the user $u$:
\begin{align}
\label{eq:softmax}
p_\theta(i \mid u) &= \frac{\exp\, s(i \mid u)}{\sum_{i' \in \mathcal{V}} \exp\, s(i' \mid u)},
\end{align}
which shares support with the prior $q$. 

The full training objective includes two terms: a next-item prediction loss, and a training-time trust region that keeps the learned distribution close to $q$,
\begin{align}
\label{eq:cg2r_objective}
\mathcal{L}_{CG2R} &= \underbrace{-\frac{1}{|\mathcal{B}|} \sum_{(u,t) \in \mathcal{B}} \log p_\theta\bigl(i^{\star}_{u,t} \mid u\bigr)}_{\mathcal{L}_{gen}: \text{ next-item prediction}} \;+\; \underbrace{\beta \, D_{\mathrm{KL}}\bigl(p_\theta \,\big\|\, q\bigr)}_{\text{trust-region penalty}},
\end{align}
where $\mathcal{B}$ is a batch of (user, position) pairs and $i^{\star}_{u,t}$ is the ground-truth next item. We set the penalty weight $\beta = 0$ in our experiments so that cluster guidance acts through the decoder alone. The guidance strength $\gamma$ is learned jointly with the model under the same objective, so the reliance on the prior is determined by training.

\begin{table}[t]
    \centering
    \caption{Statistics of datasets after removing all users and items with fewer than
five interactions.}
    \label{tab:dataset}
    \setlength{\tabcolsep}{5pt}
    \footnotesize
    \begin{tabular}{lcccc}
        \toprule
        Dataset & \#Users & \#Items & \#Interactions & Density \\
        \midrule
        MovieLens-20M & 138{,}493 & 18{,}345 & 19{,}984{,}024 & 0.7866\% \\
        Steam & 334{,}313 & 12{,}011 & 7{,}793{,}069 & 0.1941\% \\
        Google Local -- WA & 622{,}737 & 95{,}521 & 12{,}229{,}380 & 0.0206\% \\
        \bottomrule
    \end{tabular}
\end{table}

\begin{table*}[t]
\centering
\small
\caption{Overall recommendation performance. Best result per column in \textbf{bold}.}
\label{tab:main_results}
\setlength{\tabcolsep}{4pt}
\begin{tabular}{lcccccccccccc}
\toprule
\multirow{2}{*}{Method}
& \multicolumn{4}{c}{MovieLens-20M}
& \multicolumn{4}{c}{Steam}
& \multicolumn{4}{c}{Google Local -- WA} \\
\cmidrule(lr){2-5} \cmidrule(lr){6-9} \cmidrule(lr){10-13}
& R@5 & N@5 & R@10 & N@10
& R@5 & N@5 & R@10 & N@10
& R@5 & N@5 & R@10 & N@10 \\
\midrule
MF
& 0.0180 & 0.0121 & 0.0295 & 0.0158
& 0.1922 & 0.1303 & 0.2665 & 0.1542
& 0.0296 & 0.0196 & 0.0481 & 0.0255 \\

PMF
& 0.0480 & 0.0329 & 0.0718 & 0.0406
& 0.2095 & 0.1394 & 0.2962 & 0.1674
& 0.0423 & 0.0272 & 0.0696 & 0.0360 \\

NeuMF
& 0.1147 & 0.0791 & 0.1664 & 0.0958
& 0.1803 & 0.1223 & 0.2569 & 0.1470
& 0.0124 & 0.0076 & 0.0222 & 0.0108 \\

LLMRec
& 0.1310 & 0.0800 & 0.2090 & 0.1049
& 0.1945 & 0.1384 & 0.2845 & 0.1672
& 0.0301 & 0.0215 & 0.0561 & 0.0299 \\
\midrule
CG2R (ours)
& \textbf{0.2707} & \textbf{0.1864} & \textbf{0.3799} & \textbf{0.2217}
& \textbf{0.2991} & \textbf{0.2120} & \textbf{0.4067} & \textbf{0.2466}
& \textbf{0.1085} & \textbf{0.0756} & \textbf{0.1577} & \textbf{0.0915} \\
\bottomrule
\end{tabular}
\end{table*}\section{Experiments}
\subsection{Experimental Setup}
\subsubsection{Datasets}
We evaluated the proposed CG2R method on three publicly available recommendation benchmarks spanning distinct domains: MovieLens-20M~\cite{harper2015movielens} for movies, the Steam Video Game Reviews dataset~\cite{kang2018self} for video games, and the Google Local Reviews 2021 dataset~\cite{li2022uctopic,yan2022personalized} (using the \textit{Washington} subset) for local businesses such as restaurants, shops, and services. The three datasets also range differently in terms of interaction density (from 0.79\% on MovieLens to 0.02\% on Washington) and catalog size (from 12{,}011 to 95{,}521 items). To alleviate data sparsity, We adopted 5-core preprocessing~\cite{he2017neural, yue2022defending, sun2019bert4rec}, where all the users and items with fewer than five interactions were iteratively removed. The resulting dataset statistics are reported in Table~\ref{tab:dataset}.

\subsubsection{Metrics}
\label{subsec:setup_metrics}
We employed the Leave-One-Out (LOO) evaluation protocol~\cite{rendle2009bpr}. For each user, the most recent interaction was used for testing, the second most recent interaction was used for validation, and all the remaining, earlier interactions were used for training. Following the approaches in \cite{ma2019gated, sedhain2016effectiveness}, we binarized the interactions: 1) for Google Local Reviews, interactions with ratings greater than 3 are labeled positive; 2) for Steam, interactions with playtime exceeding 5.5 hours are labeled positive (yielding a positive ratio consistent with the rating-based threshold).

We report Recall@K and nDCG@K with $K \in \{5, 10\}$. Under the LOO setting, Recall@K is equivalent to Hit Rate@K, while nDCG@K further accounts for the ranking positions of the ground-truth items. During evaluation, for each user, we randomly sampled 1000 true negative items (with fixed seed across methods), and ranked them together with the single ground-truth positive item to compute Recall@K and nDCG@K.

\subsubsection{Baselines}
We compared CG2R against representative baselines spanning three recommendation paradigms.
\begin{itemize}[leftmargin=*, itemsep=2pt, topsep=3pt, parsep=0pt]
    \item \textbf{MF} and \textbf{PMF}~\cite{koren2009matrix, mnih2007probabilistic}: classical collaborative filtering, lightweight and scalable for large-scale retrieval, and still widely used in practice.
    \item \textbf{NeuMF}~\cite{he2017neural}: a neural collaborative filtering model that captures nonlinear user--item interactions and performs well on dense data.
    \item \textbf{LLMRec}~\cite{netflix2025foundation,openai2025gpt52}: an LLM-based recommender following the Netflix foundation-model approach~\cite{netflix2025foundation}. We used GPT-5.2 in high-reasoning mode here, and to mitigate LLM inference cost, we evaluated LLMRec on a 10\% stratified random sample of each test set (14,200–62,200 users depending on the dataset). 
\end{itemize}

\subsection{Experimental Results}
\subsubsection{Recommendation Performance}

Table~\ref{tab:main_results} reports Recall@K and nDCG@K on the three datasets. CG2R is the strongest method on every dataset and every metric, and the largest relative improvement occurs on Google Local Washington. Specifically, for MovieLens, the densest of the three, CG2R improves R@10 by 82\% over the best baseline (0.3799 against 0.2090 for LLMRec). For Google Local Washington, the sparsest but largest-catalog dataset, the improvement is 127\% (0.1577 against 0.0696 for PMF). Steam is the closest contest: PMF is a strong classical baseline at 0.2962, and CG2R reaches 0.4067, a 37\% improvement.

The baselines themselves separate along the density axis. NeuMF is competitive on MovieLens (R@10 0.1664) but collapses on Google Local Washington (0.0222), which is consistent with its reliance on the dense co-occurrence signal that a sparse local-business catalog does not provide. LLMRec is the strongest baseline on MovieLens, yet it falls behind PMF on both Steam and Google Local Washington, which suggests that a pretrained language prior does not make up for missing collaborative signal in every domain. In contrast, CG2R is the only method that holds up across all three regimes.

\subsubsection{Efficiency and Production Deployment}

Beyond improving ranking quality, CG2R is fast enough to serve as first-stage retrieval. In our offline benchmark experiments using a single GPU, CG2R's average end-to-end latency is 1.6--2.5\,ms per request, compared with 0.14--2.55\,ms for the reported classical and neural baselines across the three datasets. LLMRec requires 53.3--55.8\,s per request, making it less suitable for latency-sensitive applications without significant optimization.

CG2R is also robust for different user–catalog scale and sparsity regimes. Steam has a large user pool with a small catalog (334K users, 12K items), whereas the user pool and the catalog for Google Local Washington are both large (623K users, 96K items), but its interaction data are much sparser. CG2R leads the metrics on both datasets, so its performance margin does not appear to depend on a particular user/catalog scale or data density.

Two further properties of CG2R are expected to be helpful in production, though we have not evaluated them here. First,  alongside the soft trust region of Section~\ref{subsec:phase3_details} applied in this paper, the prior can also support a hard variant that restricts item scoring to the items in the top-$m$ clusters, which is an explicit $O(K)$ shortlist that avoids scoring all $N$ items. Because the prior is assembled from $K$ cluster centroids rather than $N$ item-level statistics, the guidance stays compact as the catalog grows. Second, Equation~\eqref{eq:phase2_blend} handles a cold-start user directly: the history term vanishes and the prior falls back to the population-level affinity.

\section{Conclusion and Future Work}

We presented CG2R, a three-phase recommendation framework in which a learned hierarchy of semantic clusters supplies a contextual prior that is added directly to the item scoring rule. This allows the cluster structure learned during training to influence ranking at inference. Across the three benchmark datasets evaluated, CG2R outperforms classical, neural, and LLM-based baselines on every metric we report, and its performance margin is largest on the sparsest and largest catalog.

A natural future direction is to replace the global cluster-guidance strength with a per-user or confidence-gated value, allowing the model to apply guidance more selectively.

\section{Ethical Considerations}

This study uses existing public benchmark datasets and involves no direct user intervention or new collection of personal data. As with other personalized recommenders, CG2R may reproduce biases in historical interactions, over-concentrate exposure within popular or semantically narrow clusters, and infer sensitive preferences. Deployments should therefore apply appropriate data-governance and privacy controls, audit ranking quality and exposure across relevant user and item groups, and provide mechanisms for diversity, user control, and monitoring of harmful or inappropriate recommendations.

\section{Generative AI Use}

Claude Opus models, accessed through Claude Code, assisted with implementation and data preparation, including developing preprocessing pipelines and dataset adapters and debugging training and evaluation code. The authors reviewed and validated all AI-generated code and the resulting processed data. Generative AI also assisted with language editing of the manuscript.

\bibliographystyle{ACM-Reference-Format}
\bibliography{references}


\begin{thebibliography}{24}


\ifx \showCODEN    \undefined \def \showCODEN     #1{\unskip}     \fi
\ifx \showISBNx    \undefined \def \showISBNx     #1{\unskip}     \fi
\ifx \showISBNxiii \undefined \def \showISBNxiii  #1{\unskip}     \fi
\ifx \showISSN     \undefined \def \showISSN      #1{\unskip}     \fi
\ifx \showLCCN     \undefined \def \showLCCN      #1{\unskip}     \fi
\ifx \shownote     \undefined \def \shownote      #1{#1}          \fi
\ifx \showarticletitle \undefined \def \showarticletitle #1{#1}   \fi
\ifx \showURL      \undefined \def \showURL       {\relax}        \fi
\providecommand\bibfield[2]{#2}
\providecommand\bibinfo[2]{#2}
\providecommand\natexlab[1]{#1}
\providecommand\showeprint[2][]{arXiv:#2}

\bibitem[Harper and Konstan(2015)]%
        {harper2015movielens}
\bibfield{author}{\bibinfo{person}{F.~Maxwell Harper} {and}
  \bibinfo{person}{Joseph~A. Konstan}.} \bibinfo{year}{2015}\natexlab{}.
\newblock \showarticletitle{The MovieLens Datasets: History and Context}.
\newblock \bibinfo{journal}{\emph{ACM Transactions on Interactive Intelligent
  Systems}} \bibinfo{volume}{5}, \bibinfo{number}{4} (\bibinfo{year}{2015}),
  \bibinfo{pages}{1--19}.
\newblock


\bibitem[He et~al\mbox{.}(2020)]%
        {he2020lightgcn}
\bibfield{author}{\bibinfo{person}{Xiangnan He}, \bibinfo{person}{Kuan Deng},
  \bibinfo{person}{Xiang Wang}, \bibinfo{person}{Yan Li},
  \bibinfo{person}{Yongdong Zhang}, {and} \bibinfo{person}{Meng Wang}.}
  \bibinfo{year}{2020}\natexlab{}.
\newblock \showarticletitle{{LightGCN}: Simplifying and Powering Graph
  Convolution Network for Recommendation}. In
  \bibinfo{booktitle}{\emph{Proceedings of the 43rd International ACM SIGIR
  Conference on Research and Development in Information Retrieval (SIGIR)}}.
  \bibinfo{pages}{639--648}.
\newblock


\bibitem[He et~al\mbox{.}(2017)]%
        {he2017neural}
\bibfield{author}{\bibinfo{person}{Xiangnan He}, \bibinfo{person}{Lizi Liao},
  \bibinfo{person}{Hanwang Zhang}, \bibinfo{person}{Liqiang Nie},
  \bibinfo{person}{Xia Hu}, {and} \bibinfo{person}{Tat-Seng Chua}.}
  \bibinfo{year}{2017}\natexlab{}.
\newblock \showarticletitle{Neural collaborative filtering}. In
  \bibinfo{booktitle}{\emph{Proceedings of the 26th international conference on
  world wide web}}. \bibinfo{pages}{173--182}.
\newblock


\bibitem[Hou et~al\mbox{.}(2024)]%
        {hou2024blair}
\bibfield{author}{\bibinfo{person}{Yupeng Hou}, \bibinfo{person}{Jiacheng Li},
  \bibinfo{person}{Xiangjun Fu}, \bibinfo{person}{Zhankui He},
  \bibinfo{person}{An Yan}, \bibinfo{person}{Xiusi Chen}, {and}
  \bibinfo{person}{Julian McAuley}.} \bibinfo{year}{2024}\natexlab{}.
\newblock \showarticletitle{Bridging Language and Items for Retrieval and
  Recommendation: Benchmarking {LLMs} as Semantic Encoders}.
\newblock \bibinfo{journal}{\emph{arXiv preprint arXiv:2403.03952}}
  (\bibinfo{year}{2024}).
\newblock


\bibitem[Kang and McAuley(2018)]%
        {kang2018self}
\bibfield{author}{\bibinfo{person}{Wang-Cheng Kang} {and}
  \bibinfo{person}{Julian McAuley}.} \bibinfo{year}{2018}\natexlab{}.
\newblock \showarticletitle{Self-Attentive Sequential Recommendation}. In
  \bibinfo{booktitle}{\emph{2018 IEEE 18th International Conference on Data
  Mining (ICDM)}}. IEEE, \bibinfo{pages}{197--206}.
\newblock


\bibitem[Koren et~al\mbox{.}(2009)]%
        {koren2009matrix}
\bibfield{author}{\bibinfo{person}{Yehuda Koren}, \bibinfo{person}{Robert
  Bell}, {and} \bibinfo{person}{Chris Volinsky}.}
  \bibinfo{year}{2009}\natexlab{}.
\newblock \showarticletitle{Matrix Factorization Techniques for Recommender
  Systems}.
\newblock \bibinfo{journal}{\emph{Computer}} \bibinfo{volume}{42},
  \bibinfo{number}{8} (\bibinfo{year}{2009}), \bibinfo{pages}{30--37}.
\newblock


\bibitem[Kusupati et~al\mbox{.}(2022)]%
        {kusupati2022matryoshka}
\bibfield{author}{\bibinfo{person}{Aditya Kusupati} {et~al\mbox{.}}}
  \bibinfo{year}{2022}\natexlab{}.
\newblock \showarticletitle{Matryoshka Representation Learning}. In
  \bibinfo{booktitle}{\emph{Advances in Neural Information Processing
  Systems}}, Vol.~\bibinfo{volume}{35}.
\newblock


\bibitem[Lee et~al\mbox{.}(2022)]%
        {lee2022autoregressive}
\bibfield{author}{\bibinfo{person}{Doyup Lee}, \bibinfo{person}{Chiheon Kim},
  \bibinfo{person}{Saehoon Kim}, \bibinfo{person}{Minsu Cho}, {and}
  \bibinfo{person}{Wook-Shin Han}.} \bibinfo{year}{2022}\natexlab{}.
\newblock \showarticletitle{Autoregressive image generation using residual
  quantization}. In \bibinfo{booktitle}{\emph{Proceedings of the IEEE/CVF
  conference on computer vision and pattern recognition}}.
  \bibinfo{pages}{11523--11532}.
\newblock


\bibitem[Lewis et~al\mbox{.}(2020)]%
        {lewis2020retrieval}
\bibfield{author}{\bibinfo{person}{Patrick Lewis}, \bibinfo{person}{Ethan
  Perez}, \bibinfo{person}{Aleksandra Piktus}, \bibinfo{person}{Fabio Petroni},
  \bibinfo{person}{Vladimir Karpukhin}, \bibinfo{person}{Naman Goyal},
  \bibinfo{person}{Heinrich K{\"u}ttler}, \bibinfo{person}{Mike Lewis},
  \bibinfo{person}{Wen-tau Yih}, \bibinfo{person}{Tim Rockt{\"a}schel},
  \bibinfo{person}{Sebastian Riedel}, {and} \bibinfo{person}{Douwe Kiela}.}
  \bibinfo{year}{2020}\natexlab{}.
\newblock \showarticletitle{Retrieval-Augmented Generation for
  Knowledge-Intensive {NLP} Tasks}. In \bibinfo{booktitle}{\emph{Advances in
  Neural Information Processing Systems (NeurIPS)}}.
\newblock
\shownote{arXiv:2005.11401}.
\newblock


\bibitem[Li et~al\mbox{.}(2022)]%
        {li2022uctopic}
\bibfield{author}{\bibinfo{person}{Jiacheng Li}, \bibinfo{person}{Jingbo
  Shang}, {and} \bibinfo{person}{Julian McAuley}.}
  \bibinfo{year}{2022}\natexlab{}.
\newblock \showarticletitle{UCTopic: Unsupervised Contrastive Learning for
  Phrase Representations and Topic Mining}. In
  \bibinfo{booktitle}{\emph{Proceedings of the 60th Annual Meeting of the
  Association for Computational Linguistics}}. \bibinfo{pages}{6159--6169}.
\newblock


\bibitem[Ma et~al\mbox{.}(2019)]%
        {ma2019gated}
\bibfield{author}{\bibinfo{person}{Chen Ma}, \bibinfo{person}{Peng Kang},
  \bibinfo{person}{Bin Wu}, \bibinfo{person}{Qinglong Wang}, {and}
  \bibinfo{person}{Xue Liu}.} \bibinfo{year}{2019}\natexlab{}.
\newblock \showarticletitle{Gated attentive-autoencoder for content-aware
  recommendation}. In \bibinfo{booktitle}{\emph{Proceedings of the twelfth ACM
  international conference on web search and data mining}}.
  \bibinfo{pages}{519--527}.
\newblock


\bibitem[Mnih and Salakhutdinov(2007)]%
        {mnih2007probabilistic}
\bibfield{author}{\bibinfo{person}{Andriy Mnih} {and} \bibinfo{person}{Russ~R.
  Salakhutdinov}.} \bibinfo{year}{2007}\natexlab{}.
\newblock \showarticletitle{Probabilistic Matrix Factorization}. In
  \bibinfo{booktitle}{\emph{Advances in Neural Information Processing Systems
  (NIPS)}}, Vol.~\bibinfo{volume}{20}. \bibinfo{pages}{1257--1264}.
\newblock


\bibitem[{Netflix Technology Blog}(2025)]%
        {netflix2025foundation}
\bibfield{author}{\bibinfo{person}{{Netflix Technology Blog}}.}
  \bibinfo{year}{2025}\natexlab{}.
\newblock \bibinfo{title}{Foundation Model for Personalized Recommendation}.
\newblock
\shownote{Accessed: January 2026}.
\newblock
\urldef\tempurl%
\url{https://netflixtechblog.com/foundation-model-for-personalized-recommendation-1a0bd8e02d39}
\showURL{%
\tempurl}


\bibitem[OpenAI(2025)]%
        {openai2025gpt52}
\bibfield{author}{\bibinfo{person}{OpenAI}.} \bibinfo{year}{2025}\natexlab{}.
\newblock \bibinfo{booktitle}{\emph{GPT-5.2: Technical Report}}.
\newblock \bibinfo{type}{{T}echnical {R}eport}. \bibinfo{institution}{OpenAI}.
\newblock
\shownote{Internal technical documentation, accessed January 2026}.
\newblock
\urldef\tempurl%
\url{https://openai.com/index/introducing-gpt-5-2/}
\showURL{%
\tempurl}


\bibitem[Rajput et~al\mbox{.}(2023)]%
        {rajput2023recommender}
\bibfield{author}{\bibinfo{person}{Shashank Rajput}, \bibinfo{person}{Nikhil
  Mehta}, \bibinfo{person}{Anima Singh}, \bibinfo{person}{Raghunandan~H.
  Keshavan}, \bibinfo{person}{Trung Vu}, \bibinfo{person}{Lukasz Heldt},
  \bibinfo{person}{Lichan Hong}, \bibinfo{person}{Yi Tay},
  \bibinfo{person}{Vinh~Q. Tran}, \bibinfo{person}{Jonah Samost},
  \bibinfo{person}{Maciej Kula}, \bibinfo{person}{Ed~H. Chi}, {and}
  \bibinfo{person}{Maheswaran Sathiamoorthy}.} \bibinfo{year}{2023}\natexlab{}.
\newblock \showarticletitle{Recommender Systems with Generative Retrieval}. In
  \bibinfo{booktitle}{\emph{Advances in Neural Information Processing Systems
  (NeurIPS)}}.
\newblock
\shownote{arXiv:2305.05065}.
\newblock


\bibitem[Rendle et~al\mbox{.}(2009)]%
        {rendle2009bpr}
\bibfield{author}{\bibinfo{person}{Steffen Rendle}, \bibinfo{person}{Christoph
  Freudenthaler}, \bibinfo{person}{Zeno Gantner}, {and} \bibinfo{person}{Lars
  Schmidt-Thieme}.} \bibinfo{year}{2009}\natexlab{}.
\newblock \showarticletitle{{BPR}: Bayesian Personalized Ranking from Implicit
  Feedback}. In \bibinfo{booktitle}{\emph{Proceedings of the Twenty-Fifth
  Conference on Uncertainty in Artificial Intelligence (UAI)}}.
  \bibinfo{pages}{452--461}.
\newblock


\bibitem[Schulman et~al\mbox{.}(2015)]%
        {schulman2015trust}
\bibfield{author}{\bibinfo{person}{John Schulman}, \bibinfo{person}{Sergey
  Levine}, \bibinfo{person}{Philipp Moritz}, \bibinfo{person}{Michael~I.
  Jordan}, {and} \bibinfo{person}{Pieter Abbeel}.}
  \bibinfo{year}{2015}\natexlab{}.
\newblock \showarticletitle{Trust Region Policy Optimization}. In
  \bibinfo{booktitle}{\emph{Proceedings of the 32nd International Conference on
  Machine Learning (ICML)}}. \bibinfo{pages}{1889--1897}.
\newblock


\bibitem[Schulman et~al\mbox{.}(2017)]%
        {schulman2017proximal}
\bibfield{author}{\bibinfo{person}{John Schulman}, \bibinfo{person}{Filip
  Wolski}, \bibinfo{person}{Prafulla Dhariwal}, \bibinfo{person}{Alec Radford},
  {and} \bibinfo{person}{Oleg Klimov}.} \bibinfo{year}{2017}\natexlab{}.
\newblock \showarticletitle{Proximal Policy Optimization Algorithms}.
\newblock \bibinfo{journal}{\emph{arXiv preprint arXiv:1707.06347}}
  (\bibinfo{year}{2017}).
\newblock


\bibitem[Sedhain et~al\mbox{.}(2016)]%
        {sedhain2016effectiveness}
\bibfield{author}{\bibinfo{person}{Suvash Sedhain},
  \bibinfo{person}{Aditya~Krishna Menon}, \bibinfo{person}{Scott Sanner}, {and}
  \bibinfo{person}{Darius Braziunas}.} \bibinfo{year}{2016}\natexlab{}.
\newblock \showarticletitle{On the Effectiveness of Linear Models for One-Class
  Collaborative Filtering}. In \bibinfo{booktitle}{\emph{Proceedings of the
  AAAI Conference on Artificial Intelligence}}, Vol.~\bibinfo{volume}{30}.
\newblock


\bibitem[Sun et~al\mbox{.}(2019)]%
        {sun2019bert4rec}
\bibfield{author}{\bibinfo{person}{Fei Sun}, \bibinfo{person}{Jian Liu},
  \bibinfo{person}{Jingwei Wu}, \bibinfo{person}{Changqing Pei},
  \bibinfo{person}{Xiao Lin}, \bibinfo{person}{Wenwu Ou}, {and}
  \bibinfo{person}{Pu Jiang}.} \bibinfo{year}{2019}\natexlab{}.
\newblock \showarticletitle{BERT4Rec: Sequential Recommendation with
  Bidirectional Encoder Representations from Transformers}. In
  \bibinfo{booktitle}{\emph{Proceedings of the 28th ACM International
  Conference on Information and Knowledge Management}}.
  \bibinfo{pages}{1441--1450}.
\newblock


\bibitem[Tay et~al\mbox{.}(2022)]%
        {tay2022transformer}
\bibfield{author}{\bibinfo{person}{Yi Tay}, \bibinfo{person}{Vinh~Q. Tran},
  \bibinfo{person}{Mostafa Dehghani}, \bibinfo{person}{Jianmo Ni},
  \bibinfo{person}{Dara Bahri}, \bibinfo{person}{Harsh Mehta},
  \bibinfo{person}{Zhen Qin}, \bibinfo{person}{Kai Hui}, \bibinfo{person}{Zhe
  Zhao}, \bibinfo{person}{Jai Gupta}, \bibinfo{person}{Tal Schuster},
  \bibinfo{person}{William~W. Cohen}, {and} \bibinfo{person}{Donald Metzler}.}
  \bibinfo{year}{2022}\natexlab{}.
\newblock \showarticletitle{Transformer Memory as a Differentiable Search
  Index}. In \bibinfo{booktitle}{\emph{Advances in Neural Information
  Processing Systems (NeurIPS)}}.
\newblock
\shownote{arXiv:2202.06991}.
\newblock


\bibitem[Yan et~al\mbox{.}(2022)]%
        {yan2022personalized}
\bibfield{author}{\bibinfo{person}{An Yan}, \bibinfo{person}{Zhankui He},
  \bibinfo{person}{Jiacheng Li}, \bibinfo{person}{Tianyang Zhang}, {and}
  \bibinfo{person}{Julian McAuley}.} \bibinfo{year}{2022}\natexlab{}.
\newblock \showarticletitle{Personalized Showcases: Generating Multi-Modal
  Explanations for Recommendations}.
\newblock \bibinfo{journal}{\emph{arXiv preprint arXiv:2207.00422}}
  (\bibinfo{year}{2022}).
\newblock


\bibitem[Yue et~al\mbox{.}(2022)]%
        {yue2022defending}
\bibfield{author}{\bibinfo{person}{Zhenrui Yue}, \bibinfo{person}{Huimin Zeng},
  \bibinfo{person}{Ziyi Kou}, \bibinfo{person}{Lanyu Shang}, {and}
  \bibinfo{person}{Dong Wang}.} \bibinfo{year}{2022}\natexlab{}.
\newblock \showarticletitle{Defending substitution-based profile pollution
  attacks on sequential recommenders}. In \bibinfo{booktitle}{\emph{Proceedings
  of the 16th ACM Conference on Recommender Systems}}. \bibinfo{pages}{59--70}.
\newblock


\bibitem[Zhang et~al\mbox{.}(2025)]%
        {zhang2025qwen3embedding}
\bibfield{author}{\bibinfo{person}{Yanzhao Zhang}, \bibinfo{person}{Mingxin
  Li}, \bibinfo{person}{Dingkun Long}, \bibinfo{person}{Xin Zhang},
  \bibinfo{person}{Huan Lin}, \bibinfo{person}{Baosong Yang},
  \bibinfo{person}{Pengjun Xie}, \bibinfo{person}{An Yang},
  \bibinfo{person}{Dayiheng Liu}, \bibinfo{person}{Junyang Lin},
  \bibinfo{person}{Fei Huang}, {and} \bibinfo{person}{Jingren Zhou}.}
  \bibinfo{year}{2025}\natexlab{}.
\newblock \showarticletitle{Qwen3 Embedding: Advancing Text Embedding and
  Reranking Through Foundation Models}.
\newblock \bibinfo{journal}{\emph{arXiv preprint arXiv:2506.05176}}
  (\bibinfo{year}{2025}).
\newblock


\end{thebibliography}

\appendix

\end{document}